\documentclass[11pt,twoside]{article}
\usepackage{asp2010}

\resetcounters

\begin{document}
\markboth{Appourchaux and Pall\'e}{The History of g-Mode Quest}

\title{The History of g-mode Quest}
\author{T.~Appourchaux$^1$ and P.~L.~Pall\'e$^{2,3}$
\affil{$^1$Institut d'Astrophysique Spatiale, UMR8617, Universit\'e Paris-Sud, B\^atiment\,121, 91405 Orsay Cedex, France\\
$^2$Instituto de Astrof{\'i}sica de Canarias, La Laguna, Tenerife, E-38205 Spain\\ 
$^3$Departamento de Astrof{\'i}sica, Universidad de La Laguna, La Laguna, Tenerife, E-38206 Spain}
}


\begin{abstract}
The quest for the solar gravity modes (or g modes) is key for the understanding of the structure and dynamics of the solar core.  We review the history of the solar g-mode searches which is separated in three nearly distinct eras which correspond to: the theory of g modes, the search from the ground and the search from space. The prospects of definitive solar g-mode detection are also discussed.
\end{abstract}

\section{Introduction}
Since the discovery of solar oscillations by \citet{TA_Leighton1962}, the interest in measuring the  associated solar pressure modes (p modes) and gravity modes (g modes) has been waning.  While the p modes provide extensive information on the structure and dynamics of the convection and radiative zones, the g modes promise access to the structure of the solar core.  Whereas the first detection of the global p modes was achieved by \citet{TA_AC79}, the first unambiguous detection of the g modes remains to be achieved.  The goal of this paper is to review the history of g-mode detection. 

In the first section, we will review the theory of the g modes (frequency, amplitude, lifetime).  In the second section, we will review the first ground-based attempts at detecting the g modes, while the last section will review the space-based attempts.  Each section has been bravely named after Isaac Asimov's Foundation trilogy.

\section{Foundation}
\begin{quotation}
``It is the chief characteristic of the religion of science that it works..." Isaac Asimov,  in {\it Foundation} (1942)
\end{quotation}

\begin{figure*}[!]
 \centerline{
 \includegraphics[width=0.85\textwidth]{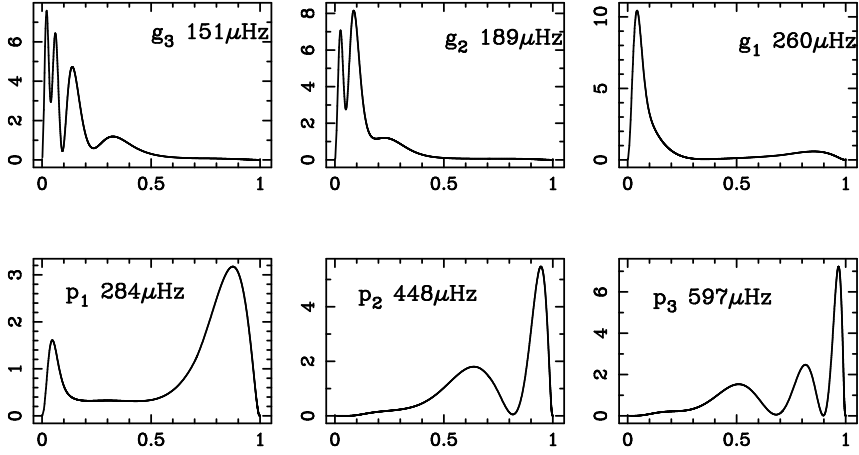}}

 \caption[]{Kinetic energy density as function of the
radius for modes g$_3$, g$_2$, g$_1$, p$_1$, p$_2$ and p$_3$ of degree
$l$=1 for a reference solar model. \citep[From][reproduced with permission \copyright\,\,ESO]{TA_JPGB2000}.}
\label{fig:TA_mixed}
\end{figure*}

For a complete review of the theory of solar oscillations, the reader may refer to \citet{TA_Appourchaux2010} and references therein.  The solar oscillations have two restoring forces, compressibility and buoyancy, characterized respectively by the sound speed $c$ and the buoyancy (Brunt -- V\"ais\"al\"a) frequency $N$.  The oscillation frequency of the gravity modes modes (g modes) is smaller that the Brunt -- V\"ais\"al\"a frequency.  As a result the amplitude of the eigenfunction of the g modes is large in the radiative zone and core of the Sun (where $N$ is not negligible), whereas the eigenfunction is large in its convective zone for the p modes (See Figure~\ref{fig:TA_mixed}).  

\begin{figure*}[!]
 \centerline{
\includegraphics[width=0.65\textwidth]{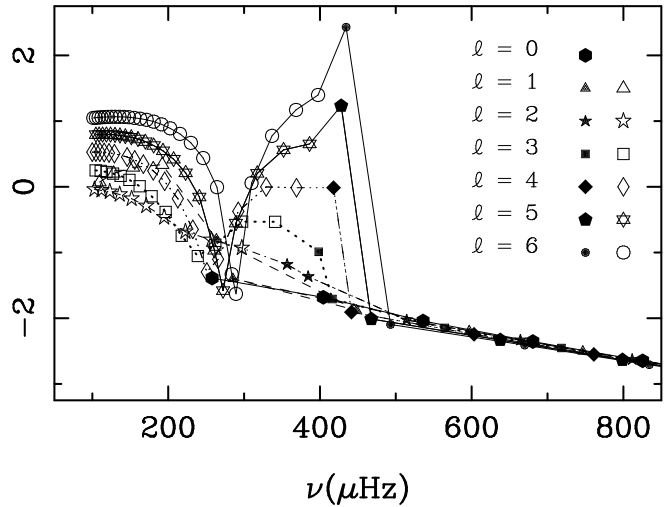}
}

 \caption[]{Logarithm of the normalized energy $\omega^2{\cal I}_{n,l}$ of low-frequency g and mixed modes (open symbols) and
low-frequency p modes (filled symbols) for a reference solar model M1
of \citet{TA_JPGB2000}, plotted as a function of frequency, for modes of
degree $l$=0, 1, 2, 3, 4, 5 and 6.  (The normalization is taken assuming that the modes have equal amplitudes  at the
photospheric level, where the temperature equals the effective
temperature level).  \citep[From][reproduced with permission \copyright\,\,ESO]{TA_JPGB2000}.}
\label{fig:TA_energie1}\end{figure*}

Figure~\ref{fig:TA_energie1} plots the normalised kinetic energy of the modes, which is proportional to the mode inertia, as a function of frequency.  Under the assumption of equipartition of the energy in the modes, the surface amplitudes would be approximately inversely proportional to the square root of the mode energy \citep[e.g.][]{TA_GB90}.  For g modes with frequencies less than 200 $\mu$Hz, the lower the degree, the smaller is the energy, and the higher is the surface amplitude.  Around 280~$\mu$Hz, modes of mixed character have smaller energies and therefore higher surface amplitudes than modes adjacent in frequency.  Note the transition from p to g modes around 450\,$\mu$Hz, and the existence of a set of modes of mixed character around 280\,$\mu$Hz, having lower energies than modes in the neighbouring frequency regions.  These latter modes, known as mixed modes, have a significant amplitude in both the interior of the Sun and in the convective zone.  These mixed modes are extremely interesting since their potential for detection is larger (they can be excited by convection) and their diagnostic potential is as good as for the proper g modes.  The detection of these mixed modes in other stars has led to great advances in the knowledge of the internal dynamics of these stars \citep{TA_Deheuvels2012}.

Low-degree g modes with frequencies less than $\nu_{n,l}\le 200\,\mu$Hz
may be described by an asymptotic relation
 \citep{TA_Vandakurov1968,TA_TASSOUL80,TA_Olver1956}. For small frequency, the second-order asymptotic expression for the period $P_{n,l}$
 ($P_{n,l}=1/\nu_{n,l}$) can be represented by
 \begin{equation}
 {P}_{n,l}\sim \overline{P}_{n,l} = {P_0\over
 L}\left(n+l /2- {1\over 4}
 +\vartheta\right)
 +\mathcal{O}\left({P_0^2\over \overline{P}_{\,n,l}}\right), 
 \label{TA_eq4J}
 \end{equation}
where $n$ is the order of the mode, $l$ is the degree of the mode, with $P_0$ given by:
 \begin{equation}
 \hbox{with}\ \ \ \ \ P_0={{2\pi^2}\over 
 \int_0^{r_{\rm cz}}{({N}/{r})\mathrm{d}r}
 },
 \end{equation}

 \noindent
Here, $\vartheta$ is a phase factor sensitive to the properties of the
layers lying below the convection zone
\citep{TA_1986A&A...165..218P,TA_Ellis86}. At the base of the convection zone, $N^2$ is assumed to vanish
proportionally to the power $p$ of the distance to the convection zone; then $\vartheta$ tends at low-frequency to a constant $- 0.5/(p+2)$.  For the standard solar model, a linear behaviour of $N^{\,2}$ may be assumed and $\vartheta$
tends to $-1/6$.  The asymptotic expression above tells us that the frequencies of g modes are related
closely to the Brunt -- V\"ais\"al\"a frequency, particularly through
$P_0$.  Typical values of $P_0$ for a
reference solar model are, $P_0 \sim 35$ to $36\,\rm min$.  The periods of g modes of a given degree $l$ are then proportional, in the
first order, to $P_0/L$, where $L=l(l+1)$.   These asymptotic properties of g modes may be exploited in attempts to
detect the modes, i.e., by searching for signatures of near-regular
patterns in period, as we shall discuss in the {\it Second foundation} section.

The lifetime and amplitude of the g modes were extensively discussed by \citet{TA_Appourchaux2010} and references therein.  The theoretical mode lifetimes are of the order of 1 million years! The main source of the g-mode damping is related to the radiative damping for $n>3$, while for lower order g modes the source is still debated \citep[See][and references therein]{TA_Appourchaux2010}.  It is assumed that the g modes are excited by turbulent convection.  The optimistic and the pessimistic values of g-mode amplitude differ by nearly 3 orders of magnitude: from 1 cm~s$^{\,-1}$ to 10$^{\,-3}$ cm~s$^{\,-1}$ for $l=1$!  This large variation is primarily due to the way that the turbulent eddies are time-correlated, i.e. with a Gaussian or a Lorentzian profile  \citep{TA_Appourchaux2010}.

The theoretical framework set for the p-mode oscillation was put to test with the detection of the p modes by \citet{TA_Deubner75}.  From this point of view, the measured ($k,\omega$) diagram of  solar oscillations confirmed the theoretical work of \citet{TA_Ulrich1970}.  The success achieved with the p modes was expected to be useful for the detection of g modes.  In other words, the g-mode theoretical predictions described in this section would be put to test by the observations.  The detection of low-frequency modes (not to say g modes) was to mark the start of the next era.

\section{Foundation and Empire}
\begin{quotation}
``It is the invariable lesson to humanity that distance in time, and in space as well, lends focus. It is not recorded, incidentally, that the lesson has ever been permanently learned." Isaac Asimov in {\it Foundation and Empire} (1945)
\end{quotation}

The first detection of low-frequency oscillations is attributed to \citet{TA_Severnyi1976} who claimed having detected a 160-min oscillation using 9 days of solar radial velocity derived from a modified differential Babcock solar magnetograph, contemporaneously confirmed by \citep{TA_JB76} using a full-disk resonance spectrometer.  \citet{TA_Severnyi1976}  identified this oscillation as being the $l=2$ g$_{11}$ mode of 2 m s$^{-1}$ amplitude.  This oscillation was later detected by several other instruments: differential Babcock solar magnetograph \citep{TA_Scherrer1979}, full-disk resonance spectrometer \citep{TA_GG80}.  The mode sensitivity of the differential magnetograph to solar oscillations was the highest for $l=3,4$ modes unlike the full-disk spectrometers that are not sensitive to these modes \citep{TA_Kosovichev1986}.  Since the mode sensitivity was different, a common source different from the Sun could explain the oscillation: the Earth's rotation.  The main problem of that oscillation is that it was very close to the ninth harmonics of the day which at 104.09 $\mu$Hz.  Last but not least, the excitation of that single mode was not easily explained; for instance \citet{TA_Kosovichev1986} listed 13 different explanations as excitation process.  Back in the 80's; it was then realized that in order to confirm the detection of any low-frequency oscillation, there would be a need for long observation either performed from ground or (even more likely) from space.

The motivation for g-mode detection was so high that it lead to the study of a space mission, the Dual Spectral Irradiance and Solar Constant Orbiter (DISCO), which showed  what could be achieved were the g modes to be detected \citep{TA_Bonnet81,TA_Balogh81}.  The prospects of g-mode detection were pushed forward by \citet{TA_PDPS83} who claimed the detection of  tens of g modes, also using a modified Babcock solar magnetograph.  In the mid-80s, the enthusiasm for a space mission was extremely active as several reports were written for the promotion of space-based observations \citep{TA_Noyes1984,TA_Marquedant1985}, and also publicized by the famous Snowmass conference \citep{TA_Ulrich1984}.  Following the non-selection of DISCO, a new mission was proposed: the Solar and Heliospheric Observatory (SoHO).  It was selected by the European Space Agency (ESA) in November 1982 for an assessment study.

The promotion for a space mission then became  a transatlantic affair with support from the whole European and American helioseismic community.  
The first author of this paper can vividly remember that when presenting my first conference paper \citep{TA_Appourchaux84} having been asked by a high-ranking official from CNES (Centre National d'Etudes Spatiales, the French space agency) {\it to show that the g modes cannot be detected from the ground}.   More important and less personal, the possible excitation of the 160-minute oscillation by the Geminga object was advanced not in a scientific journal but in the main stream French newspaper {\it Le Monde} (October 12, 1983).  This finding was relayed in the scientific journal {\it Nature} (October 20, 1983) as a {\it News and Views}.  In this same issue of {\it Nature}, it was reported that the SoHO mission was selected by ESA for a Phase-A study.  {\it Nature} asked a selection committee member whether [...] {\it the announcement influenced} [them], and the answer was: {\it It helped but was by no means decisive, SoHO would have been approved anyway}.  The possible excitation of the 160-min oscillation was in fact neither confirmed \citep{TA_Anderson1984} nor even theoretically possible \citep{TA_Bonazzola1984,TA_Carroll1984,TA_Fabian1984,TA_Kosovichev1984,TA_Kuhn1984,TA_Deruelle1984}.  The SoHO Phase-A was eventually carried out under an ESA and NASA (National Aeronautics and Space Administration) collaboration for a selection as a mission in 1986 \citep{TA_Antonucci85}.  

Space missions were not the only solution for having long-duration observations, high data fill and low noise.  There were several initiatives that led to the creation of ground-based networks such as the Global Oscillation Network Group (GONG) funded in 1984 by the National Science Foundation, USA  \citep{TA_Harvey1996}; the Birmingham Solar Oscillation Network (BiSON) operating in July 1981 \citep{TA_YE91}; or the International Research on the Interior of the Sun (IRIS) network operating in July 1984 \citep{TA_Fossat91}.

In that same decade, the fate of the 160-min oscillation was definitely settled as it was shown that it was an artefact due to spurious solar velocity induced by atmospheric extinction \citep{TA_YE89}.  In the following years, the g-mode identification by \citet{TA_PDPS83} could never be reproduced showing again the dire needs for space-based observations.

With the selection of the SoHO payload in 1988, the Global Oscillations at Low Frequency (GOLF) instrument \citep{TA_Gabriel95}, the Michelson Doppler Imager for the Solar Oscillation Investigation \citep[MDI/SOI;][]{TA_PS95} and the Variability of Irradiance and Gravity Oscillations instrument \citep[VIRGO;][]{TA_CF97} together with the aforementioned ground-based networks consistuted a formidable armada of instruments that was the start of {\it The Empire} of helioseismology.  In those days, the detection of g modes was thought to be a only matter of time.

\section{Second Foundation}

\begin{quotation}
``To any who know the star field well from one certain reference point, stars are as individual as people. Jump ten parsecs, however, and not even your own Sun is recognizable."  Isaac Asimov in {\it Second Foundation} (1948)
\end{quotation}

The prospects of g-mode detection was put to rest from the selection of SoHO to shortly before its launch.  It was revived when \citet{TA_DT95} claimed having detected g modes in particle data in the {\it Ulysses} and {\it Voyager} missions.  The detection was based upon the application of a patent widely used in cellular telephone \citep{TA_Lindberg91}.  The idea was to detect if there were a signal above the noise at a known-in-advance frequency.  Then \citet{TA_DT95} used theoretical g-mode frequencies and previously measured p-mode frequencies for detecting these modes.  The finding of \citet{TA_DT95} were not confirmed by \citet{TA_Riley1996,TA_DD99, TA_GHPR98}.  To date their detection has indeed not been confirmed.

Following the launch of SoHO on December 2, 1995, the perspective of g-mode detection started exciting times.  \citet{TA_Palle1998a} put a definite end to the existence of the 160-min oscillation as GOLF was not detecting anything at all.  Two years after the launch, it was realized that the helioseismic instruments of SoHO could not individually detect the g modes.  In 1997, the first author of this paper decided to follow the legacy of the late Philippe Delache (known as the ambassador of g modes) and started a working group composed of member of the teams of MDI/SOI, VIRGO and BiSON.  The group was named after Gaston III Phoebus (Count of Foix) famous for having written {\it The Book of Hunt} (1388), a book very appropriate for hounds!  The Phoebus group held 5 workshops from 1997 to 2002 which led to several publications of attempts at detecting the g modes \citep{TA_Chaplin2002,TA_Wachter2003} amongst which one provided an upper limit to g-mode amplitude of 1\,cm~s$^{\,-1}$ \citep{TA_TA2000}.  The many different techniques devised for detecting the g modes can be found in \citet{TA_Appourchaux2010} and references therein.  

On the other hand, the GOLF team was separately trying to find the g modes in their own data.  They reported having detected an $l=1$ g$_{\,1}$ mode at 284.7 $\mu$Hz but still provided an upper limit to the g-mode amplitude of 0.6 cm s$^{\,-1}$ \citep{TA_Gabriel02}; and the detection of several candidates \citep{TA_TC2004} of amplitude 0.6\,cm s$^{\,-1}$ at around 220 $\mu$Hz, akin to an $l=2$ g$_{\,3}$ mode.  At the time of writing, none of these finding are confirmed.

The Phoebus group was then enlarged with the addition of members of the GOLF team.  The Phoebus group then held three additional workshops from 2005 to 2007 that led to additional attempts at g-mode detections especially with the introduction of Bayesian-type of detection  \citep{TA_Appourchaux2008b,TA_AMB2007,TA_AMB2010}.  During that time, using the asymptotic property laid out in Equation~(\ref{TA_eq4J}), \citet{TA_RAG2007} reported a collective detection of g-mode signature.  They found a peak in the periodogram of the periodogram of the GOLF data performed in the range [60,140] $\mu$Hz.  Using this collective detection of these peaks, they reported individual detection of the peaks using the {\it collapsogramme} technique with a typical signal-to-noise ratio of 2 for $l=1$ g$_{\,8}$ mode and a splitting of about 890\,nHz, resulting in an amplitude of 0.1 cm~s$^{\,-1}$ \citep[See also][for an application of the collapsogramme to the detection of low order p modes]{TA_Salabert2009}.  At the time of writing, the collective detection has not been confirmed using techniques different from that of the periodogram of the periodogram \citep{TA_AMB2010}.

The Phoebus activities were stopped in 2010 with the publication of a review paper which stated that {\it there [was] indeed a consensus
amongst the authors of this review that there [was] currently no undisputed detection of solar g  modes.} \citep{TA_Appourchaux2010}.  Unfortunately, the first author still believes that this consensus holds in\,2013.

\begin{figure}[!]
\begin{center}
\includegraphics[width=0.45\textwidth,angle=90]{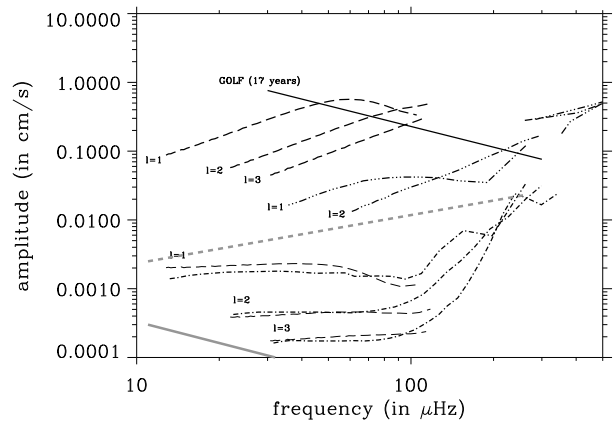}
\end{center}
\caption[]{Estimated amplitudes of stochastically excited 
g modes of low degree plotted against frequency $\nu$. The estimates
are rms surface values for singlet modes (single values of $n$, $l$ and $m$),
joined by lines: triple-dot-dashed for \citet{TA_Gough85}, dot-dashed for \citet{TA_Kumar96} 
and dashed for \citet{TA_KB2009}. The continuous line is an 
estimate of the 10\% limit from 17 years GOLF data \citep{TA_RAG2007}.  The thick grey lines are two estimates from \citet{TA_Shiode2013} for two eddy sizes of 100\% (dashed line) and of 30\% (solid line) of the pressure scale height.  For proper comparison, the effect of the spatial instrumental filter must be included, i.e. visibilities \citep{TA_Appourchaux2010}.
}
\label{fig:TA_amp_dog_kumar_kevin}
\end{figure}

\section{Perspective}

\begin{quotation}
``The observer influences the events he observes by the mere act of observing them or by being there to observe them" Isaac Asimov in {\it Foundation's Edge} (1982)
\end{quotation}

At the time of the conference in Tucson, several of my colleagues provided me with recent results of their own g-mode search.  Scherrer and Larson (2013, private communication) tried to find a coincident peak in the $l=1$ MDI periodogram of the periodogram with that of \citet{TA_RAG2007} but could not find any evidence common peaks.  Fossat\,(these proceedings) computed a time series of the mean location of the p-mode envelope and its associated power spectrum.  He then used the periodogram of the periodogram in the range [25,40] $\mu$Hz hoping to detect an $l=1$ comb, thanks to the asymptotic properties given by Equation~(\ref{TA_eq4J}).  Then using the likely frequency location of the modes, he co-added 8 possible g-mode spectra in order to enhance them that provided a splitting of 439 nHz.  Needless to say that these two results show that there is still a lot of interest in the detection of the g modes.

Very recently, \citet{TA_Shiode2013} studied how g modes could be excited by convection in stars more massive than the Sun, stars having convective cores.  They used their model to derive solar g-mode amplitudes which are indeed in the ball park of previous estimates (See Figure~\ref{fig:TA_amp_dog_kumar_kevin} for a comparison of the various estimates).

The four independent estimates of g-mode amplitude given in Figure~\ref{fig:TA_amp_dog_kumar_kevin} compared to the GOLF 17-year limit shows that it is very likely that the solar g modes have not been detected yet.  If we were to trust the most optimistic value of \citet{TA_KB2009}, we would have been able to detect the g modes
with a signal-to-noise ratio of 2, a value commonly achieved in asteroseismology \citep[See][as an example]{TA_Appourchaux2008}.  It is then very likely that the g-mode amplitude are at least 10 times lower than the values provided for $l=1$ by \citet{TA_KB2009}.  

What do you need in order to detect the solar g modes?  There has been two possibilities: to dig deeper the current database of ground- and space-based observations, and to reduce the solar atmospheric noise with new instruments.  As for the first solution, this has been done over the past 30 years and will likely continue for several decades to come.  As for the second solution, there has been two ideas that have not been yet implemented.  

The first idea is developed under the concept of the Global Oscillations of Low-Degree modes \citep[GOLD;][]{TA_STC2012} which aims at measuring solar radial velocities at different heights in the atmosphere, thereby hoping to reduce the solar noise using the cross spectrum technique of \citet{TA_RG1999}.  Unfortunately, it has been shown that even if the coherence between two signals is zero, the average of the cross spectrum will indeed be zero but the variance of the cross spectrum will not follow this behaviour \citep{TA_TA2007}.  Assuming that the g modes have very long lifetime then the advantage of the GOLD instrument is only for low-degree solar p modes with a lifetime shorter than the observing time \citep{TA_TA2007}.

The second idea is to directly measure the gravitational perturbation generated by the g modes, that is the strain or the deformation of spacetime.  This is what measures the ASTROD (Astrodynamical Space Test of Relativity using Optical Devices) mission proposed to ESA's Cosmic Vision \citep{TA_Appourchaux2009b}.  The strain sensitivity of ASTROD would allow 
to detect solar g modes with the lowest predicted amplitudes of Figure~\ref{fig:TA_amp_dog_kumar_kevin}, a sensitivity at least 2 order of magnitude better than the LISA (Laser Interferometer Space Antenna) mission of ESA \citep{TA_Appourchaux2009b}.  At the time of writing, LISA has not been selected as a mission but LISA pathfinder, a mission for testing in flight the concept of low-frequency gravitational wave detection, will be launched in 2015.  As for the ASTROD mission, there is no programmatic window since it is not known when it will fly as no space agency has it in their long-term plans.

Finally, there remains the possibility of testing the current model of solar g modes amplitudes using the data from {\it Kepler}.  If we could detect g modes in heavier stars, hoping thereby that the amplitudes are higher in more massive stars, then the excitation model could then be validated.  WIth such a validation, it would be possible either to continue the search with the actual data or to devise more sensitive measurement techniques.  With such a positive mind, we can then anticipate that we are to reach the {\it Third foundation} of g-mode detection.

\acknowledgements TA would like to thank three important players who contributed to his helioseismic professional career: Jacques-Emile Blamont who introduced him to the world of space missions, of solar seismology and of tenacity; Pierre Connes who taught him the basics of proper and sound instrument design and  last but not least David M. Rust whose confidence, openness and daring vision made him TA what he is today. 

\bibliography{Appourchaux}
\end{document}